\documentclass[letter]{aa}         

\usepackage{natbib}
\usepackage{graphicx}
\usepackage{txfonts}
\usepackage{hyperref}
\usepackage{xcolor}

\newcommand{\sgs}{SMCSGS-FS\,69}

\DeclareRobustCommand{\ion}[2]{%
	\relax\ifmmode
	\ifx\testbx\f@series
	{\mathbf{#1\,\mathsc{#2}}}\else
	{\mathrm{#1\,\mathsc{#2}}}\fi
	\else\textup{#1\,{\mdseries\textsc{#2}}}%
	\fi}
\def\degr{\hbox{$^\circ$}}

\newcommand{\hei}{ \ion{He}{i}\ }
\newcommand{\heii}{ \ion{He}{ii}\ }
\newcommand{\ha}{H$\alpha$\ }
\newcommand{\hb}{H$\beta$\ }
\newcommand{\hg}{H$\gamma$\ }
\newcommand{\hd}{H$\delta$\ }

\def\apjl{ApJ}%
\def\apjs{ApJS}%
\def\ao{Appl.~Opt.}%
\def\aap{A\&A}%
\begin{document}

  \title{A partially stripped massive star in a Be binary at low metallicity%
  \thanks{Based on observations at the European Southern Observatory Very
Large Telescope programs 086.D-0167(A) and 0103.D-0921(A)
}}
  \subtitle{A missing link towards Be X-ray binaries and double neutron star mergers}
  \author{V. Ramachandran\inst{1} \and J. Klencki\inst{2,3}  \and A. A. C. Sander\inst{1}  \and D. Pauli\inst{4}
          \and T. Shenar\inst{5}    \and L. M. Oskinova\inst{4} \and W.-R. Hamann\inst{4}           }

   \institute{Zentrum f{\"u}r Astronomie der Universit{\"a}t Heidelberg,
Astronomisches Rechen-Institut, M{\"o}nchhofstr. 12-14, 69120 Heidelberg\\
              \email{vramachandran@uni-heidelberg.de}
\and European Southern Observatory, Karl-Schwarzschild-Strasse 2, 85748, Garching bei München, Germany  
\and Max Planck Institute for Astrophysics, Karl-Schwarzschild-Strasse 1, 85748, Garching bei München, Germany
\and Institut f{\"u}r Physik und Astronomie, Universit{\"a}t Potsdam, Karl-Liebknecht-Str. 24/25, D-14476 Potsdam, Germany  
\and Anton Pannekoek Institute for Astronomy, University of Amsterdam, 1090 GE Amsterdam, The Netherlands
             }

\date{}

\abstract{Standard binary evolutionary models predict a significant population of core helium-burning stars that lost their hydrogen-rich envelope after mass transfer via Roche-lobe overflow. However, there is a scarcity of observations of such stripped stars in the intermediate-mass regime ($\sim 1.5- 8\,M_\odot$), which  are thought to be prominent progenitors of SN Ib/c. Especially at low metallicity, a significant fraction of these stars are expected to be only partially stripped, retaining a significant amount of hydrogen on their surfaces.  For the first time, we discovered a partially stripped massive star in a binary with a Be-type companion located in the Small Magellanic Cloud (SMC) using a detailed spectroscopic analysis. 
The stripped-star nature of the primary is revealed by the extreme CNO abundance pattern and very high luminosity-to-mass ratio, which suggest that the primary is likely shell-hydrogen burning. 
Our target \sgs\ is the most luminous and most massive system among the known stripped star + Be binaries, with $M_{\mathrm{stripped}}\sim 3\,M_\odot$ and $M_{\mathrm{Be}} \sim17\,M_\odot$. Binary evolutionary tracks suggest an initial mass of  $M_{\mathrm{ini}} \gtrsim12\,M_\odot$ for the stripped star and predict it to be in a transition phase towards a hot compact He star, which will eventually produce a stripped-envelope supernova. Our target marks the first representative of an as-yet-missing evolutionary stage in the formation pathway of Be X-ray binaries and double neutron star mergers.}


\maketitle


\section{Introduction}

Massive stars are frequently found in binary or multiple systems where components are in close proximity to one another, making the interaction between the two stars inevitable as the primary grows and evolves \citep{Sana2012,Sana2014,Moe2017}. The components will be subjected to substantial interactions involving the transfer of mass and angular momentum,  which will have profound effects on the fundamental parameters and final fates of both stars. Such interactions frequently lead to the stripping of the primary's envelope \citep{Kippenhahn1967,Paczynski1967}, which can produce hot and compact He-core stars with a thin layer of hydrogen on top \citep[see, e.g.,][]{Yoon2010,Yoon2017,Claeys2011}. Depending on their initial masses, the stripped-envelope primaries would have spectral characteristics ranging from hot subdwarfs to Wolf-Rayet (WR) stars \citep[e.g.,][]{Paczynski1967,Vanbeveren1991,Eldridge2008,gotberg_2017_ionizing}. Secondaries, on the other hand, would evolve into rapidly rotating stars \citep[e.g.,][]{deMink2013,Renzo2021}, which can have disk emission features similar to Be stars \citep{Pols1991,Shao2014,Bodensteiner+2020MS,Hastings2021}.

Interestingly, stripped-envelope stars with masses between low-mass subdwarfs and classical WR stars are rarely observed. This intermediate-mass regime $\sim 1.5- 8\,M_\odot$ at solar metallicity \citep{gotberg_2017_ionizing} gets wider up to $\lesssim 17\,M_\odot$ at SMC metallicity \citep[Z=0.2\,$Z_{\odot}$; ][]{Shenar2020}. Moreover, the intermediate-mass stripped-envelope stars (hereafter stripped stars) are predicted to be a long-lived core He-burning phase. They are considered to be the progenitors of Ib/c supernovae and major sources of far-UV ionizing flux \citep{gotberg_2017_ionizing}. The only known intermediate-mass hot He star is the qWR star HD 45166 \citep{Groh2008}, whereas other hot and compact stripped star candidates in the Galaxy are in the subdwarf mass range of $<1.5\,M_{\odot}$ \citep{Wang2021,Schootemeijer2018,Gilkis2023}. For HD\,45166, a new study yields a strong magnetic field, meaning that this star likely does not follow standard binary evolution \citep{Shenar2023submitted}.

In recent literature, there is growing evidence for stripped stars, but many of them are partially stripped OB-type giants of a few solar radii, often with a significant residual H-layer ($X_\mathrm{H}>50\%$) on their surface. These include originally proposed X-ray quiet black hole + Be binaries, such as LB1 and HR6819 \citep{Liu2019,Rivinius2020}, which have later been  disputed \citep[e.g.,][]{Abdul-Masih2020,Bodensteiner2020HR6819,Shenar2020lb1,El-Badry2021} or were revealed to be partially stripped star + Be binaries \citep{Frost2022}. Another similar object is HD\,15124, which is currently undergoing mass transfer \citep{El-Badry2022b}. All these objects are in the Milky Way and have masses $\lesssim1.5\,M_{\odot}$. In contrast, \cite{Irrgang2022} reported $\gamma$\,Columbae to be a partially stripped pulsating core ($\sim\!4-5\,M_{\odot}$) of a massive star that has a spectral appearance similar to that of a B subgiant but with altered surface abundances. However, there is no evidence of a mass-accreted secondary in this system. In the Large Magellanic Cloud (LMC), notable systems are NGC1850\,BH1 \citep{El-Badry2022,Saracino2023} and NGC\,2004\#115 \citep{Lennon2022,El-Badry2022NGC2004}, which are speculated to contain a black hole or low-mass stripped star. Growing observational evidence for these partially stripped stars compared to the apparent absence of fully stripped He stars in the intermediate-mass regime raises questions about our understanding of binary evolution.
  
Recent evolutionary models by \cite{Klencki2022} suggest that binary evolution at low metallicity favors partial envelope stripping and slow mass transfer, leading to a large population of partially stripped donors. Due to their predicted surface properties, these systems are likely hiding among OB binaries as apparent main sequence (MS) or supergiant stars \citep[e.g.,][]{Pauli2022}. Identifying and characterizing such systems at low metallicity would yield sharp constraints on binary evolution and, in turn, on the origin of gravitational wave sources, ultra-luminous X-ray sources, stripped-envelope supernovae, and ionization in the early Universe.
In this paper we report the first observational evidence of a partially stripped star + Be binary at low metallicity.


\section{Observations}

   \begin{figure}
   \centering
   \includegraphics[width=8cm]{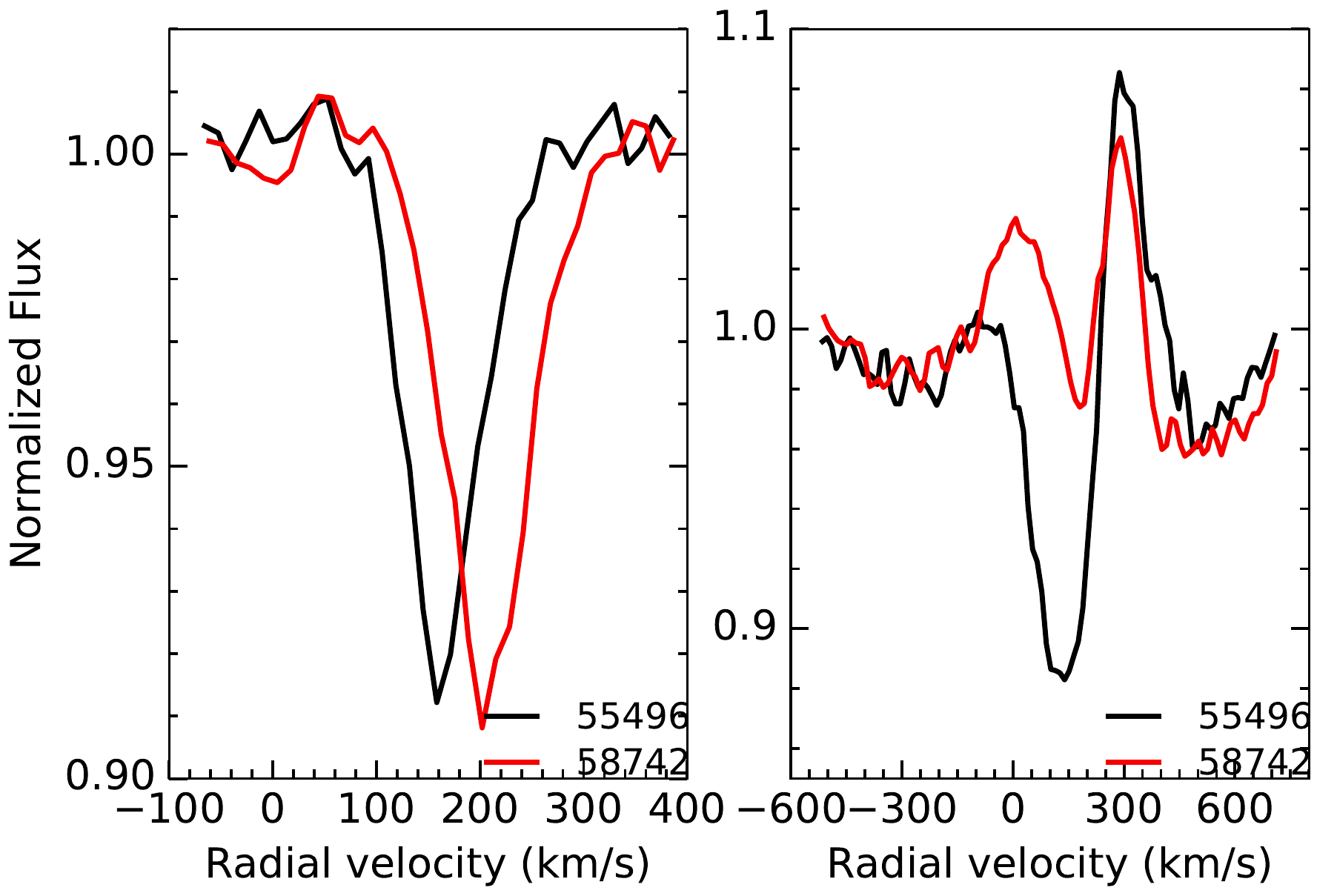}
      \caption{Radial velocity variation in \ion{Si}{iii}$\lambda$4553 (left) and H$\beta$ (right). The colors indicate the two observation epochs in Julian dates (see legend). 
              }
         \label{RV}
   \end{figure}

  \begin{figure}
   \centering
   \includegraphics[width=8.5cm]{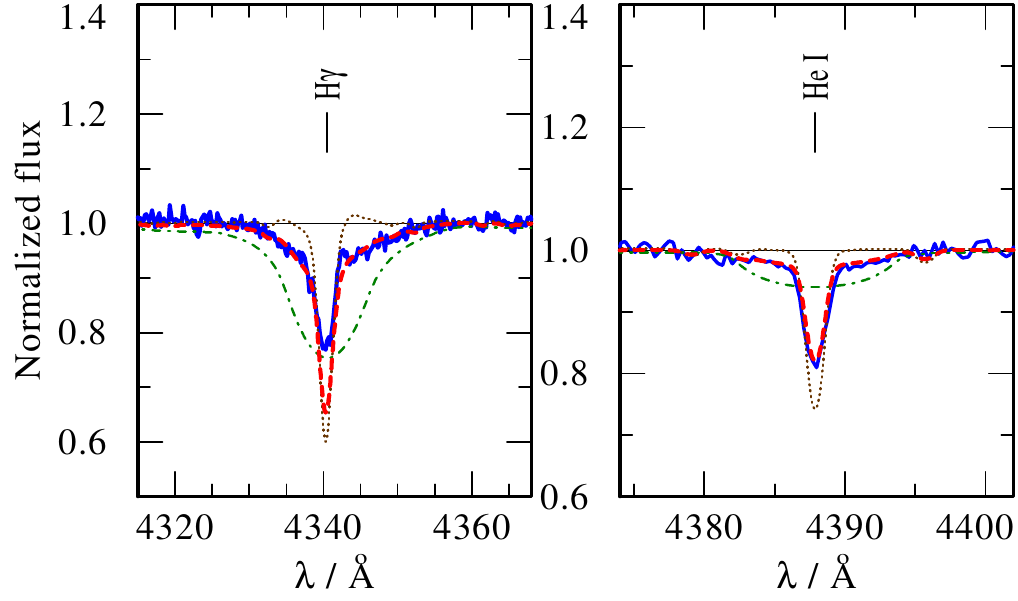}
      \caption{Observed spectra of \sgs\, (blue solid lines) displaying narrow and broad components.  The composite
model (dashed red) is the weighted sum of the stripped star primary (dotted brown)
and the rapidly rotating Be star secondary (dashed green) model spectra with effective temperatures of
24\,kK and 28\,kK, respectively. The weaker absorption core in H$\gamma$ likely results from filled-in disk emission.
              }
         \label{bin}
   \end{figure}

In a previous study \citep{ramachandran2019} we analyzed the optical 
spectra of OB stars in the Wing of the SMC taken in 2010 with the Fiber Large Array Multi-Element Spectrograph (FLAMES) on the ESO Very Large Telescope (VLT).  Three of the standard settings of the Giraffe spectrograph LR02 (resolving power $R$\,=\,6000, 3960--4567\,\AA), LR03 ($R$\,=\,7500, 4501--5071\,\AA), and HR15N ($R$\,=\,19200, 6442--6817\,\AA) were used for this survey.  Details of the observation, data reduction, and extraction of the spectra are described in \citet{ramachandran2019}. As a spectroscopic follow-up, we collected additional epochs in 2019 for most of this sample. 
 
In this work we carefully inspected the spectra of Be stars and other fast rotators in this sample and discovered that \sgs\, shows significant radial velocity (RV) variations up to 45\,km\,s$^{-1}$ (see Fig.\,\ref{RV} and more details in Appendix\,\ref{RVestimate}). 
Based on a single epoch optical spectrum the star was initially classified as B0.5\,(II)e \citep{ramachandran2019}. With a detailed inspection of the spectra, we found that \sgs\, is a double-line spectroscopic binary, consisting of both broad- and narrow-line components (Fig.\,\ref{bin}), indicating  that it is a potential post-mass-transfer binary. In addition, high-resolution H-band (1.51-1.70 $\mu$m) spectra from the Apache Point Observatory Galactic Evolution Experiment (APOGEE) survey \citep{Majewski2017} are available in the archive, which we used to study the Be companion. Furthermore, our target was observed with Gaia and has listed proper motions and parallaxes. The radial velocities from the optical spectra and the Gaia proper motions agree well with that of SMC Wing. The negative values for the Gaia parallaxes support that this object is not a foreground Galactic star.

In addition to the spectra, we used various photometric data (from UV to infrared) from VizieR to construct the spectral energy distribution (SED). We also utilized data from the Transiting Exoplanet Survey Satellite (TESS) for this system. We extracted the light curves and found that the variability cannot be consistent with the orbital period, but rather indicates rotational modulations induced by the Be star (see Appendix\,\ref{tess} for details).

\section{Analysis}

In the optical range  the overall spectrum of \sgs\ resembles an early B-type supergiant except for the following a) disk emission features in \ha and \hb; b) the presence of extended broad wings and narrow absorption in \hg and \hd;  c) a combination of strong narrow absorption and weak broad components in multiple \hei lines; and d) the strength of CNO absorption lines different from typical supergiant spectra. These features imply that the observed spectrum is a composite of a slowly rotating partially stripped B supergiant-like star and a fast-rotating MS star with disk emission (Fig.\,\ref{bin}). All the metal lines are narrow, suggesting they are mainly from the stripped star. Although \ha is in emission, it is not clear whether this is entirely from the Be secondary or if there is a contribution from the stripped primary. Despite the low S/N, all the Brackett series lines in the APOGEE spectra (Fig.\,\ref{fig:IR}) display broad disk emission with multiple peaks, which are mostly  from the secondary Be star.
 
We performed the spectral analysis of \sgs\ using the PoWR model atmosphere code  \citep[see][for more details]{Graefener2002,Hamann2004,Sander2015}.
Initially, we chose PoWR SMC grid models \citep{Hainich2019} as a starting point for the investigation and further computed additional models with tailored parameters for the primary and secondary. The full spectral and SED fit is shown in Fig.\,\ref{fig:optfit}. 
 
  \begin{figure}
   \centering
   \includegraphics[scale=0.75, angle =90]{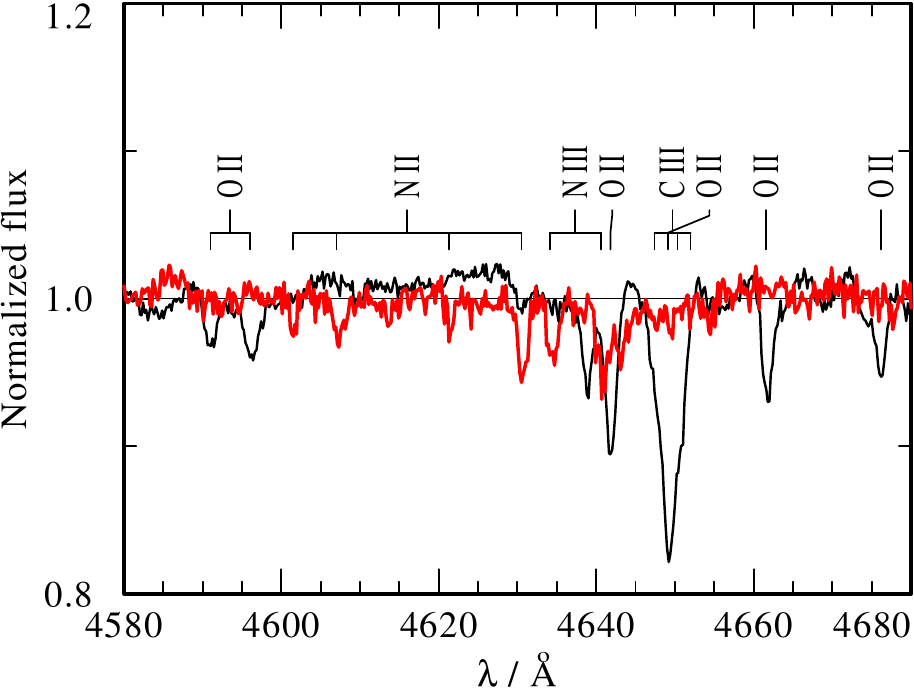}
      \caption{Comparison of CNO lines in the spectra of \sgs\, (red) and an  SMC B supergiant (AV\,242) of similar $T_\mathrm{eff}$ (black). The spectra demonstrate that carbon and oxygen are substantially reduced, while nitrogen is highly enriched in \sgs.
              }
         \label{spec_comp}
   \end{figure}
   
In the spectral fitting method, we started with the analysis of the primary since it has a major contribution in the optical. First, we estimated the projected rotation velocity ($\varv\,\sin i$) of narrow-lined primary from metal absorption lines. We used a combined Fourier transform (FT) and goodness-of-fit (GOF) analysis employing the \texttt{iacob-broad} tool \citep{Simon-diaz2014}. We applied this method to several metal lines and found the overall mean to deduce $\varv\sin i$ and macro-turbulent ($\varv_{\mathrm{mac}}$) velocities. Subsequently, these values, along with instrumental broadening, were used to convolve the model spectra to match the observations. 

The main diagnostic we use to constrain the temperature of the primary is the He and Si ionization balance based on \ion{He}{i}\,/\,\ion{He}{ii} and \ion{Si}{iii}\,/\,\ion{Si}{iv} line ratios. The pressure-broadened wings of the Balmer lines are the primary diagnostics for the surface gravity. We considered H$\gamma$ and H$\delta$ since they are less impacted by wind and disk emissions. However, in our case the H$\gamma$ and H$\delta$ have contributions from both primary and secondary. Thus we simultaneously adjusted the luminosity ratios and surface gravities in the primary and the secondary models to match the observations. Since the ionization balance also reacts to gravity, we simultaneously re-adjusted $T_\ast$ and $\log\,g_\ast$ to achieve a good fit to the observed spectra. The final uncertainty in the primary parameters reflects the overall fit quality and is limited by the model grid resolution. Constraining the secondary parameters is challenging, but we find that assuming cooler temperatures results in stronger broader components in the metal and \hei lines, whereas hotter temperatures lead to pronounced \heii lines in the composite spectra. To account for the broad lines from the secondary, we had to use a very high  $\varv\,\sin i$.

\begin{table}[!hbt]
	\caption{Fundamental parameters and abundances derived for SMCSGS-FS\,69 using spectroscopic analysis.}
	\label{table:parameters}
	\centering
	\renewcommand{\arraystretch}{1.6}
	\begin{tabular}{lcc}
		\hline 
		\hline
		\vspace{0.1cm}
&	Stripped star                                     &     Be star           \\
		\hline 
		$T_{\ast}$ (kK)                                & $24^{+2}_{-1}$       &  $28^{+2}_{-3}$ \\
  $T_{2/3}$ (kK)                                 & $21^{+2}_{-1}$ &  $28^{+2}_{-3}$ \\
   
		$\log g_\ast$ (cm\,s$^{-2}$)    & $2.65^{+0.2}_{-0.1}$   &   $3.7^{+0.2}_{-0.2}$   \\
  $\log g_{2/3}$ (cm\,s$^{-2}$)    & $2.4^{+0.2}_{-0.1}$    &    $3.7^{+0.2}_{-0.2}$   \\
  flux $f / f_{\mathrm{tot}}$\,(V) &0.65 & 0.35 \\
		$\log L$ ($L_\odot$)                           & $4.7^{+0.1}_{-0.1}$  &   $4.7^{+0.15}_{-0.15}$    \\ 
		$R_\ast$ ($R_\odot$)                           & $13^{+1}_{-1.5}$    &    $8.7^{+2}_{-1.5}$      \\
   $R_{2/3}$  ($R_\odot$)                               & $17^{+1.5}_{-2}$  &   $8.7^{+2}_{-1.5}$    \\
		$\varv \sin i$ (km\,s$^{-1}$)       & $50^{+10}_{-10}$   &  $400^{+100}_{-100}$      \\
		$\varv_{\mathrm{mac}}$ (km\,s$^{-1}$)         & $20^{+20}_{-10}$   &    50 (fixed)   \\
		$\xi$ (km\,s$^{-1}$)            & $12^{+3}_{-2}$     &      $14$  (fixed)   \\
		$X_{\rm H}$ (mass fr.)                         & $0.59^{+0.1}_{-0.1}$  &   0.737\tablefootmark{$\ast$}  \\
		$X_{\rm He}$ (mass fr.)                         & $0.40^{+0.1}_{-0.1}$  &   0.26\tablefootmark{$\ast$}   \\
		$X_{\rm C}/10^{-5}$ (mass fr.)                 & $\lesssim 1$       &   21\tablefootmark{$\ast$}     \\
		$X_{\rm N}/10^{-5}$ (mass fr.)                 & $120^{+20}_{-20}$  & 3\tablefootmark{$\ast$}          \\
		$X_{\rm O}/10^{-5}$ (mass fr.)                 & $\lesssim7$      &   113\tablefootmark{$\ast$}      \\
		$X_{\rm Si}/10^{-5}$ (mass fr.)                & $11^{+2}_{-2}$    &  13\tablefootmark{$\ast$}       \\
		$X_{\rm Mg}/10^{-5}$ (mass fr.)                & $19^{+3}_{-3}$ &  10\tablefootmark{$\ast$}  \\                    
		$E_{\mathrm{B-V}}$ (mag)                                & $0.13^{+0.03}_{-0.03}$ &    \\
		$M_\mathrm{spec}$ ($M_\odot$)                  & $2.8^{+1.5}_{-0.8}$    &  $17^{+9}_{-7}$      \\
		$\log\,Q_{\mathrm H}$ (s$^{-1}$)   & 47.3  &  47.5   \\
		$\log\,Q_{\mathrm {He\,\textsc{ii}}}$ (s$^{-1}$)   & 32.9  &  37.5   \\
		\hline
	\end{tabular}
	\tablefoot{ \tablefoottext{$\ast$} {Abundances of the Be star are adopted from \citet{Trundle2007} which corresponds to typical values for OB stars in the SMC}
 } 
 \vspace{-0.2cm}
\end{table}

Although we initially used grid models computed with typical SMC abundances \citep{Trundle2007}, they do not reproduce satisfactorily the observed CNO lines. To match the observed strength of CNO absorption lines, we had to drastically increase N and decrease C and O in the primary model. Most B supergiants in the SMC do not show such pronounced abundance variations. This is illustrated in Fig\,\ref{spec_comp}, where we compare the optical spectra of \sgs\, to that of an SMC B supergiant.
The N abundance is determined by analyzing multiple \ion{N}{ii} and \ion{N}{iii} lines, while the CO abundance is just an upper limit as most of the CO lines are either very faint or within the noise. In addition to the CNO abundances, we varied the H mass fraction ($X_{\rm H}$) in the primary models between 0.5 and 0.73 and found that slightly He-enriched models better represent the observations.
 The remaining elements either had their abundance values scaled to one-fifth of solar or adopted typical SMC abundances derived from OB stars \citep{Trundle2007}. We also checked for the overall broadening of metal lines by varying the micro-turbulence $\xi$ in the range of 10-20\,km\,s$^{-1}$.
  
Since a UV spectrum is not available for this object, we can only constrain the wind parameters ($\dot{M}$ and $\varv_\infty$) from H$\alpha$. The H$\alpha$ profile shows a single but asymmetric emission peak (Fig.\,\ref{fig:optfitha}). If this emission is only contributed by the primary stripped star, it can be modeled as a result of a strong and slow stellar wind ($\log \dot{M}\!\approx\!-6.2$ and $\varv_{\infty} \approx600$\,km\,s$^{-1}$), as illustrated in Fig.\,\ref{fig:optfitha} (left). However, since infrared spectra (Fig.\,\ref{fig:IR}) clearly showcase multi-peak disk emission features, we cannot exclude disk emission components in H$\alpha$. We can alternatively reproduce the asymmetric \ha profile with a combination of strong disk emission from the Be star and a weak absorption component ($\log \dot{M}\!=\!-7.2$) from the stripped star. Consequently, the wind mass-loss rate could be lower or higher depending on the Be disk emission strength. To precisely constrain the wind parameters, it is necessary to obtain UV spectra and to disentangle the components using multi-epoch optical spectra.

We determine the luminosity $L$ and color excess $E_{\rm B-V} $ by fitting the composite model SED to the photometry (first panel of Fig.\,\ref{fig:optfit}). The model flux is diluted with the SMC Wing distance modulus of 18.7\,mag \citep{Cignoni2009}. By fitting the normalized spectra, we are able to place constraints on the luminosity ratio, thus the SED fitting by composite model yields both primary and secondary luminosities. Both components were found to have the same luminosity, even though the primary stripped star contributes approximately 60-65\%  of the light in the optical range.

\section{Results and discussion}

Our spectroscopic analysis reveals that while the estimated temperature and gravity of the primary are in the range of B supergiants, the luminosity, radius, and consequently mass are considerably lower. The spectroscopic mass of the primary ($2-4.3 M_{\odot}$) is strikingly low compared to what is expected for an early B supergiant. Notably, the partially stripped star's \heii flux contribution is much smaller than  expected from fully stripped stars \citep[e.g.,][]{gotberg_2017_ionizing}, which is mainly due to the much lower temperature.
The spectroscopic mass of the Be star secondary is less constrained ($10-26 M_{\odot}$) but consistent with a MS star. A complete overview of the derived parameters of both the primary and the secondary is given in Table\,\ref{table:parameters}.

 \begin{figure}
   \centering
   \includegraphics[width=8cm]{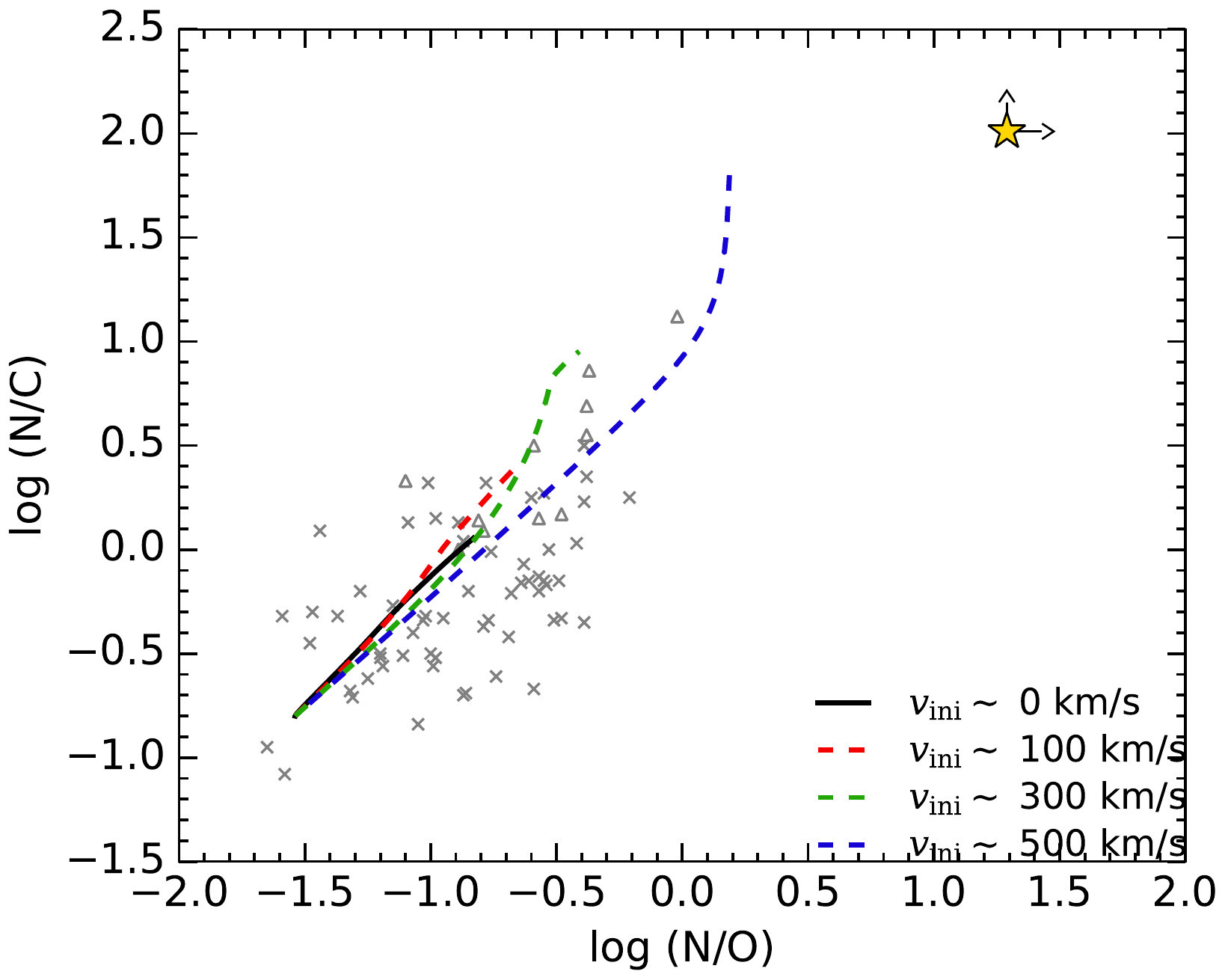}
      \caption{N/C vs. N/O abundances of the stripped star (yellow star) compared to SMC B stars (triangles for  giants and crosses for  main sequence) from \cite{Hunter2009}. The dashed lines are tracks from \citet{Brott2011} for the SMC with different initial rotations for $M_{\mathrm{ini}}=12$. The location of the stripped star at the top right corner demonstrates that its surface CNO pattern is too extreme to be explained by standard stellar evolution. }
         \label{NC}
\end{figure}

\begin{figure}
   \centering
   \includegraphics[width=7.5cm]{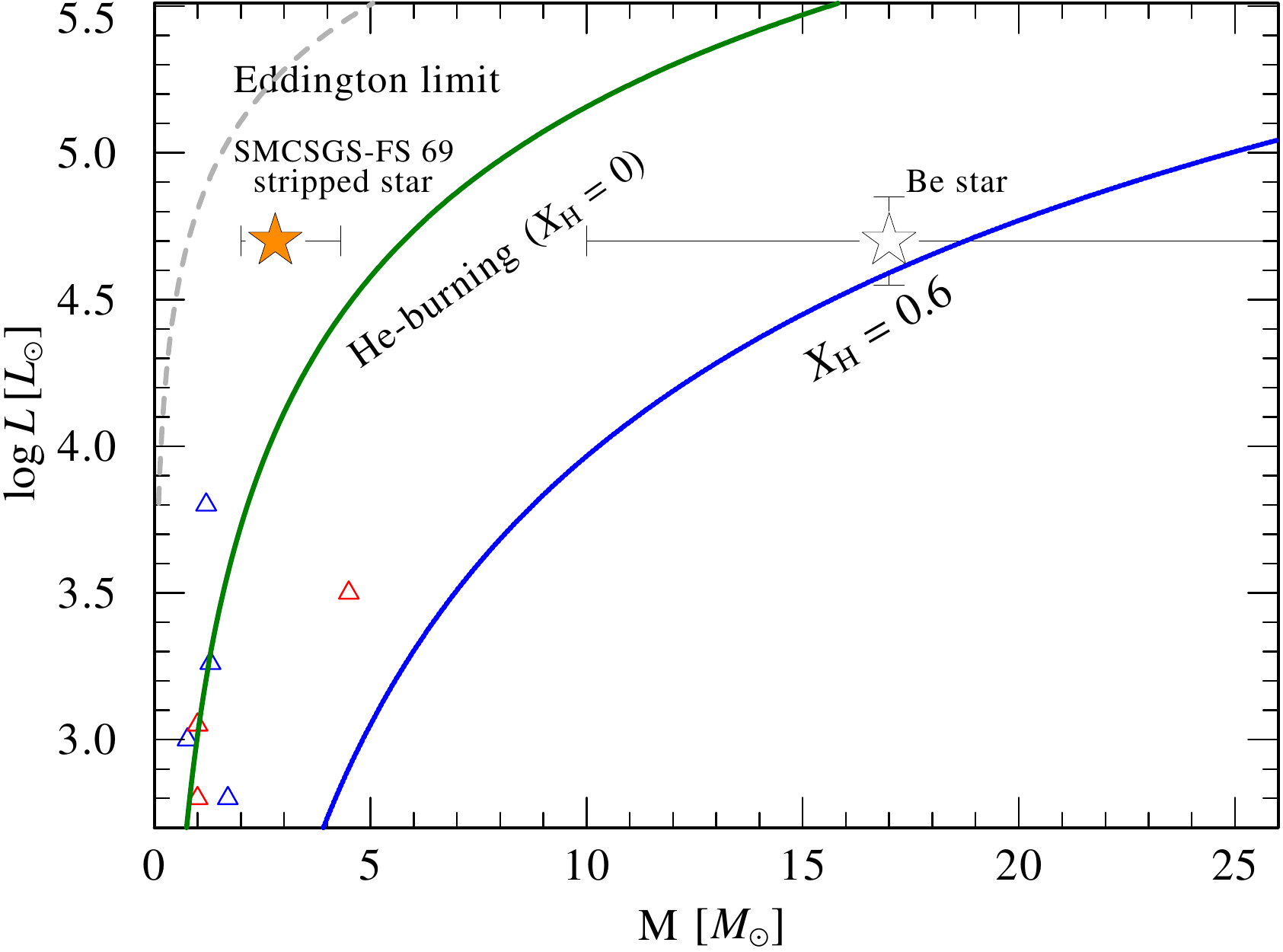}
      \caption{Location of the \sgs\, components compared to mass-luminosity relations for pure He stars (green line) and early MS stars ($X_{\rm H}$ = 0.6, blue line), from \cite{Graefener2011}.  The position of the partially stripped primary is shown by a filled star, while the Be secondary is shown by an open star along with their respective uncertainties. For comparison, Galactic hot subdwarfs in binaries (blue triangles) and partially stripped stars (red triangles) are indicated.
      The stripped primary is located in the upper left corner, which is a characteristic of a star that has lost most of its envelope via mass transfer.      }
         \label{ML}
\end{figure}

In Fig.\,\ref{NC}, we illustrate that, unlike other SMC B stars including supergiants, our primary star shows an extreme surface abundance pattern that cannot be explained via rotational mixing \citep[e.g., as calculated in the tracks by][]{Brott2011}. The derived nitrogen abundance is a factor of 4 higher than for typical SMC B supergiants and ten times that for average B stars \citep{Dufton2005,Hunter2007}.   
On the other hand, carbon and oxygen are depleted by more than a factor of $\sim$\,15 compared to typical SMC stars.
While silicon abundances are in agreement, a slight enrichment in magnesium is detected. A comparison of the derived surface abundances with typical B stars and B supergiants in the SMC is given in the Appendix Table\,\ref{table:abundance}. 

The unusual abundance pattern can be explained by the removal of external layers via mass transfer, exposing the inner CNO-processed layers of the star. This would be in line with the exceptionally high luminosity-to-mass ratio of the primary, which we illustrate in Fig.\,\ref{ML} and compare it to relations of pure He stars and  (early) MS stars. Notably, its luminosity ($\log L/L_\odot\!\sim\!4.7$) is much higher than that of a pure He star of the same mass ($\log L/L_\odot\!\sim\!4.0$), which suggests that most of the luminosity is produced in a leftover envelope layer via H-shell burning. Our primary is located in the upper left corner of Fig.\,\ref{ML}, close to the Eddington limit, which is characteristic for stars that have lost a significant amount of mass via mass transfer. Altogether, these clues indicate a partially stripped star nature for the primary. 

\begin{figure}
   \centering
   \includegraphics[width=8cm]{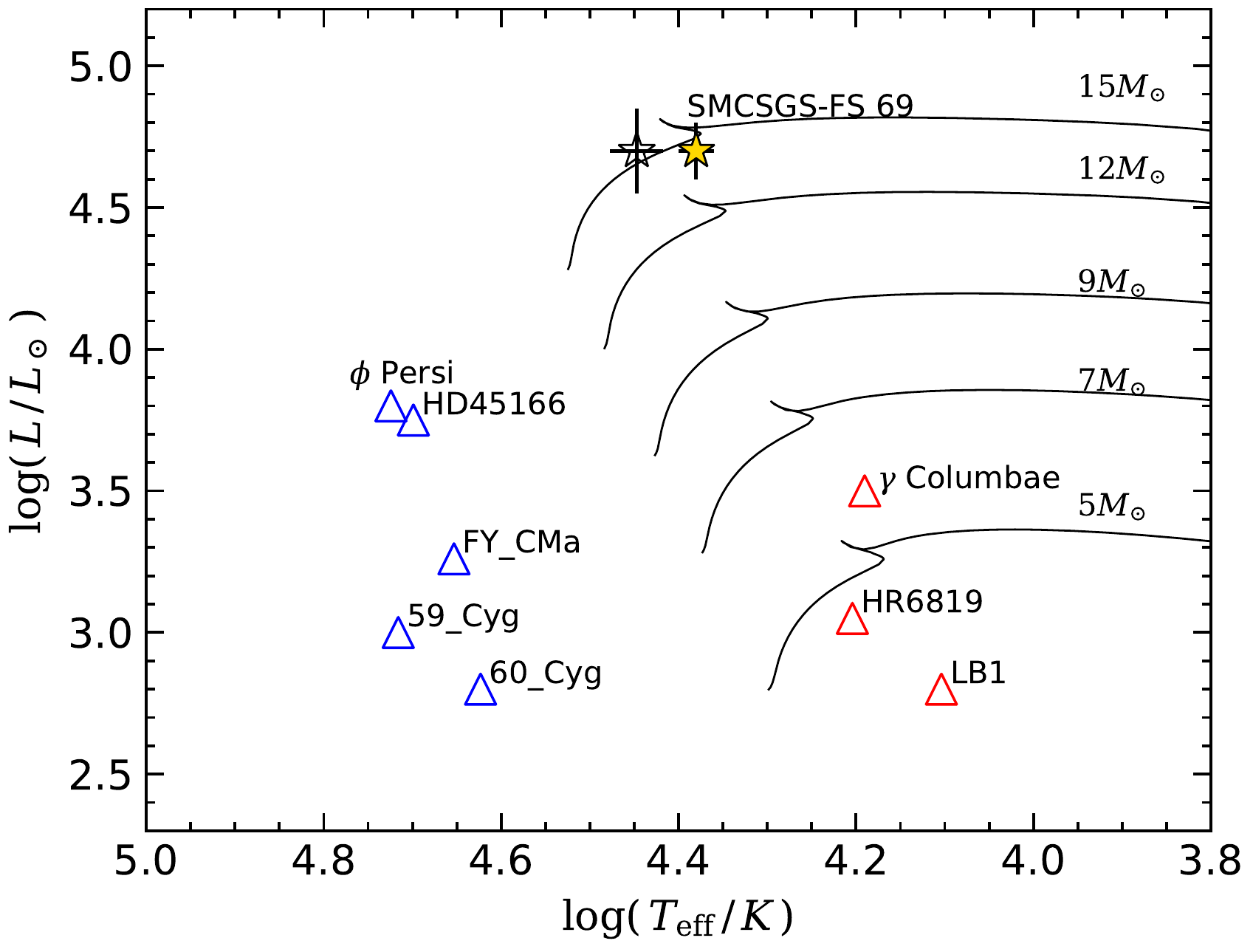}
      \caption{Location of \sgs\, on the HR diagram compared to other stripped star candidates from the literature. Bonn SMC tracks are plotted for comparison \citep{Brott2011}. The symbols are the same as in Fig.\,\ref{ML}}
         \label{hrdsingle}
\end{figure}

In the Hertzsprung–Russell (HR) diagram in Fig.\,\ref{hrdsingle} we  compare \sgs\, with the other stripped star candidates reported in the literature, including fully and partially stripped objects. The newly discovered \sgs\, stands out as the most luminous (and massive) of all and is the first one detected at SMC metallicity. The HRD in Fig.\,\ref{hrdsingle} shows again that its mass is much lower than that of a MS star with a similar luminosity ($\sim 15 M_{\odot}$).

\begin{figure}
   \centering
   \includegraphics[width=\columnwidth]{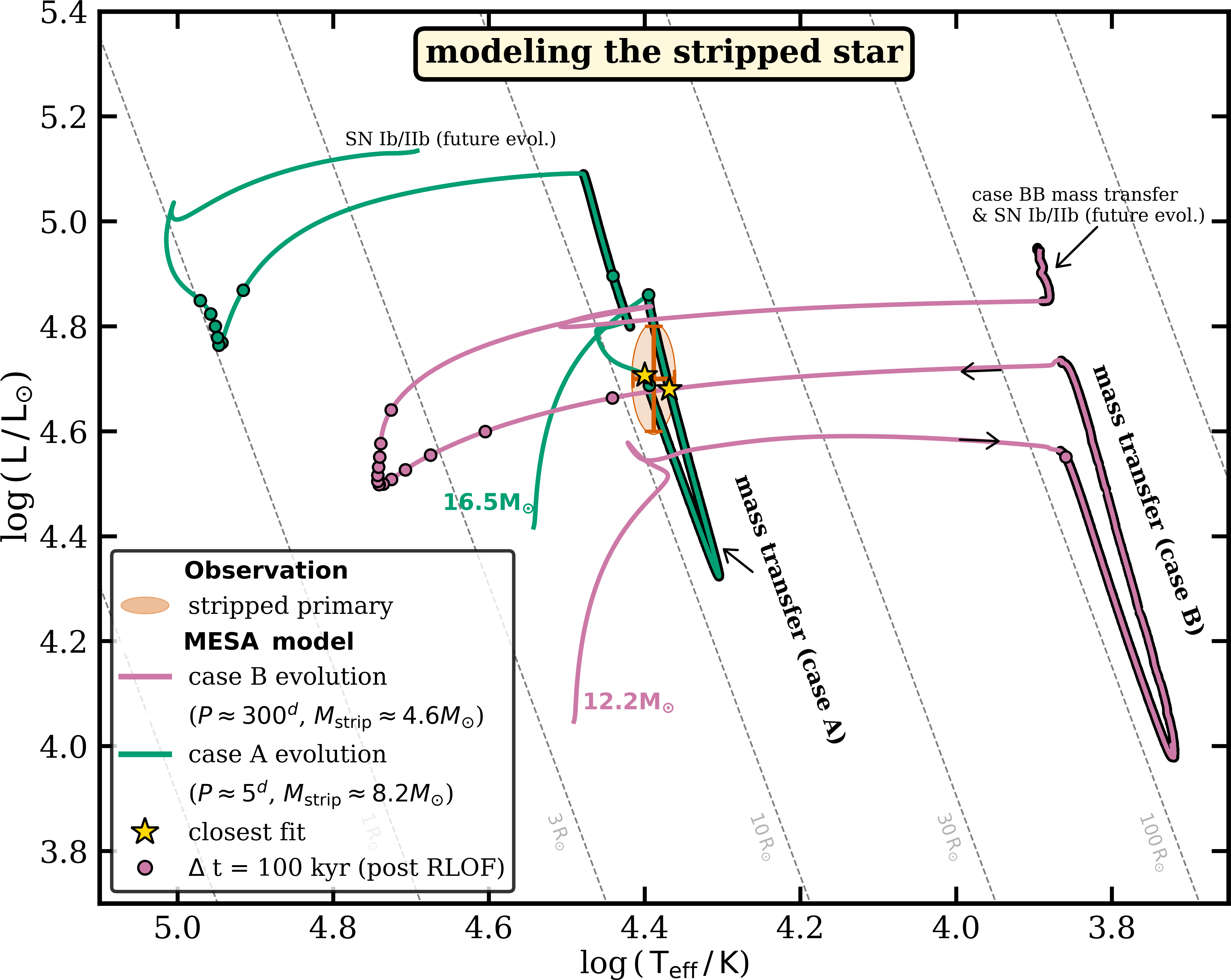}
      \caption{Two potential binary evolution pathways leading to the formation of a stripped star in \sgs. The Case B mass transfer evolution roughly matches most of the derived surface properties, though it notably requires $\gtrsim 40\%$ accretion efficiency to reproduce the secondary (Fig.\,\ref{fig:hrd_caseB}). The Case A mass transfer evolution does not have that requirement, but it overpredicts the current mass of the stripped star.}
         \label{hrd_bin}
\end{figure}

To interpret the evolutionary status of the system, we employed the stellar evolution code MESA \citep{Paxton2015,Paxton2018,Paxton2019} to compute a binary model for \sgs\, (see Appendix\,\ref{sec:app_mesa} for details).
We found several solutions that are able to reproduce several of the observed parameters. Notably, based on a grid of $\sim2300$ binary evolution models, we find that the luminosity of the stripped primary can only be reproduced by models with the current mass stripped star$\gtrsim3 M_{\odot}$.
The closest match can be obtained with a Case B model  in which the primary has an initial mass of $12.2 M_{\odot}$ and is stripped via Case B mass transfer to produce a partially stripped star of $4.6 M_{\odot}$ (with a He core of $3.6 M_{\odot}$).
This is illustrated by the magenta line in Figures \,\ref{hrd_bin} and \ref{fig:param}. The Case B model (marginally) matches most of the properties derived in our spectral analysis, except for the surface O abundance (which is a factor of $\sim$ 3 too large). In order to reproduce the luminosity of the Be star companion, we find that at least $\sim 40\%$ of the mass transferred during a Case B interaction needs to be accreted by the secondary. (see the example of an $11.7 M_{\odot}$ secondary accreting $3 M_{\odot}$ in Fig.\,\ref{fig:hrd_caseB}). This may be in tension with models where the accretion efficiency is regulated via surface rotation of the accretor, which generally predicts negligible accretion during Case B evolution \citep{Sen2022,Pauli2022}, but may reach higher values depending on the assumed angular momentum budget \citep[$\sim 30\%$ in the model by][]{Renzo2021}. We find that an alternative solution in which no mass accretion is strictly required could possibly be obtained if the stripped star originates from a more massive primary ($16.5 M_{\odot}$) that interacts already during the MS (i.e., Case A mass transfer), and we observe the stripped product while it is still core-H burning. This scenario is illustrated by the green line in Fig.\,\ref{hrd_bin} and in Fig.\,\ref{fig:hrd_caseA}. The Case A model matches the observed surface abundances and rotation velocity of the stripped star very well, but is inconsistent regarding the current mass (by $\gtrsim 4 M_{\odot}$) and surface gravity (by $\gtrsim 0.45$ dex). While the determination of the surface gravity is affected by the rotationally broadened lines of secondary, the discrepancy in $\log g$ is so large that we consider the Case A scenario (under the current evolutionary calculation scheme) to be less likely. We estimate the relative rate of the two scenarios to be comparable in the population of the SMC within a factor of $\sim$2 (see Appendix\,\ref{sec:app_mesa}).

Regardless of whether the stripped star in \sgs\ is a product of Case A or Case B mass transfer evolution, its pre-interaction mass of $\gtrsim 12 M_{\odot}$ guarantees that it sits firmly in the mass regime for the formation of neutron stars (NSs), making it the first such stripped star found to date. Our binary evolution models suggest that in 1$-$1.5 Myr the primary will explode as a type IIb/Ib supernova (with $\sim 0.02 M_{\odot}$ of H left at the surface) and form a NS. If the system remains bound, it will later evolve into a Be X-ray binary. The favored Case B model tentatively suggests a long orbital period of hundreds of days at this stage. Such wide Be X-ray binaries are thought to be the direct progenitors of common-envelope events, leading to double NS mergers \citep{Tauris2017,vignagomez2020,Klencki2021}. The properties of Be X-ray binaries in the SMC seem to point toward moderate accretion efficiencies in their prior mass transfer evolution \citep{Vinciguerra2020,Igoshev2021}, in agreement with our Case B evolutionary model for \sgs. These considerations emphasize the significance of \sgs\ as a newly discovered intermediate evolutionary stage in the formation pathway of Be X-ray binaries and double NS mergers.

The discovery of a massive stripped star + Be binary in a transition phase allows binary evolution to be constrained and hints at the existence of more such binaries at low metallicity. Moreover, \sgs\ can act as a template for identifying hidden systems in typical OB populations. These partially stripped transition stages are more luminous due to their H-shell burning and are visible in the optical due to the cooler surface temperatures. 
Even though the transition phase is short-lived ($\lesssim10\%$ He burning lifetime), binary evolution models predict tens or more of similar objects to be hiding among the known OB star population of the SMC (Klencki et al., in prep), motivating a large-scale search for binary-interaction products in this low-metallicity environment.

The growing population of partially stripped stars, including \sgs,
further raises the question of why we discover systems during a short transition stage, but so far do not observe intermediate-mass stripped stars in their hot, compact stage. Evolutionary models usually predict that stripped stars settle at hotter temperatures, but with lower luminosities (see, e.g., Fig.\,\ref{hrd_bin}). It is presently unclear whether these objects are just very hard to observe or this stage might not be regularly reached by binary evolution, contrary to current predictions. The presence or absence of such a compact stripped-star population will have a severe impact on population synthesis predictions, for example, due to the different evolutionary fates and ionizing fluxes.

\begin{acknowledgements}
We thank the anonymous referee for useful suggestions. 
We would like to thank Ylva G\"{o}tberg and Cole Johnston for helpful discussions.
VR and AACS are supported by the Deutsche Forschungsgemeinschaft (DFG - German Research Foundation) in the form of an Emmy Noether Research Group -- Project-ID 445674056 (SA4064/1-1, PI Sander) and funding from the Federal Ministry of Education and Research (BMBF) and the Baden-Württemberg Ministry of Science as part of the Excellence Strategy of the German Federal and State Governments. JK acknowledges support from the ESO Fellowship. DP acknowledges financial support by the Deutsches Zentrum f\"ur Luft und Raumfahrt (DLR) grant FKZ 50OR2005.
 TS acknowledges support from the European Union's Horizon 2020 under the Marie Skłodowska-Curie grant agreement No 101024605. This research was supported by the International Space Science
Institute (ISSI) in Bern, through ISSI International Team project \#512
"Multiwavelength view on massive stars in the era of multimessenger astronomy".  
   
\end{acknowledgements}

\bibliographystyle{aa} 
\bibliography{ref} 

\appendix
\section{Radial velocity}
\label{RVestimate}
Four epochs of spectra are available for \sgs\,; three of them are taken within one day and one with a gap of nine years.
To estimate the RVs, we used a template to cross-correlate selected spectral lines (\ion{Si}{iii}$\lambda 4553$) or full spectra, and then fit a parabola to the cross-correlation function's maximum region (e.g., Zucker 2003). Initial RV estimates were always calculated using one of the observations as the template.
We found a maximum $\delta_{\mathrm{RV}}\approx 45$\,km\,s$^{-1}$ for spectra taken in 2010 and 2019, and $\delta_{\mathrm{RV}}\lesssim 10$\,km\,s$^{-1}$ for spectra taken within one day.

\section{TESS light curve}
\label{tess}
\begin{figure}
   \centering
   \includegraphics[width=8cm,trim={0 2cm 1cm 2cm}]{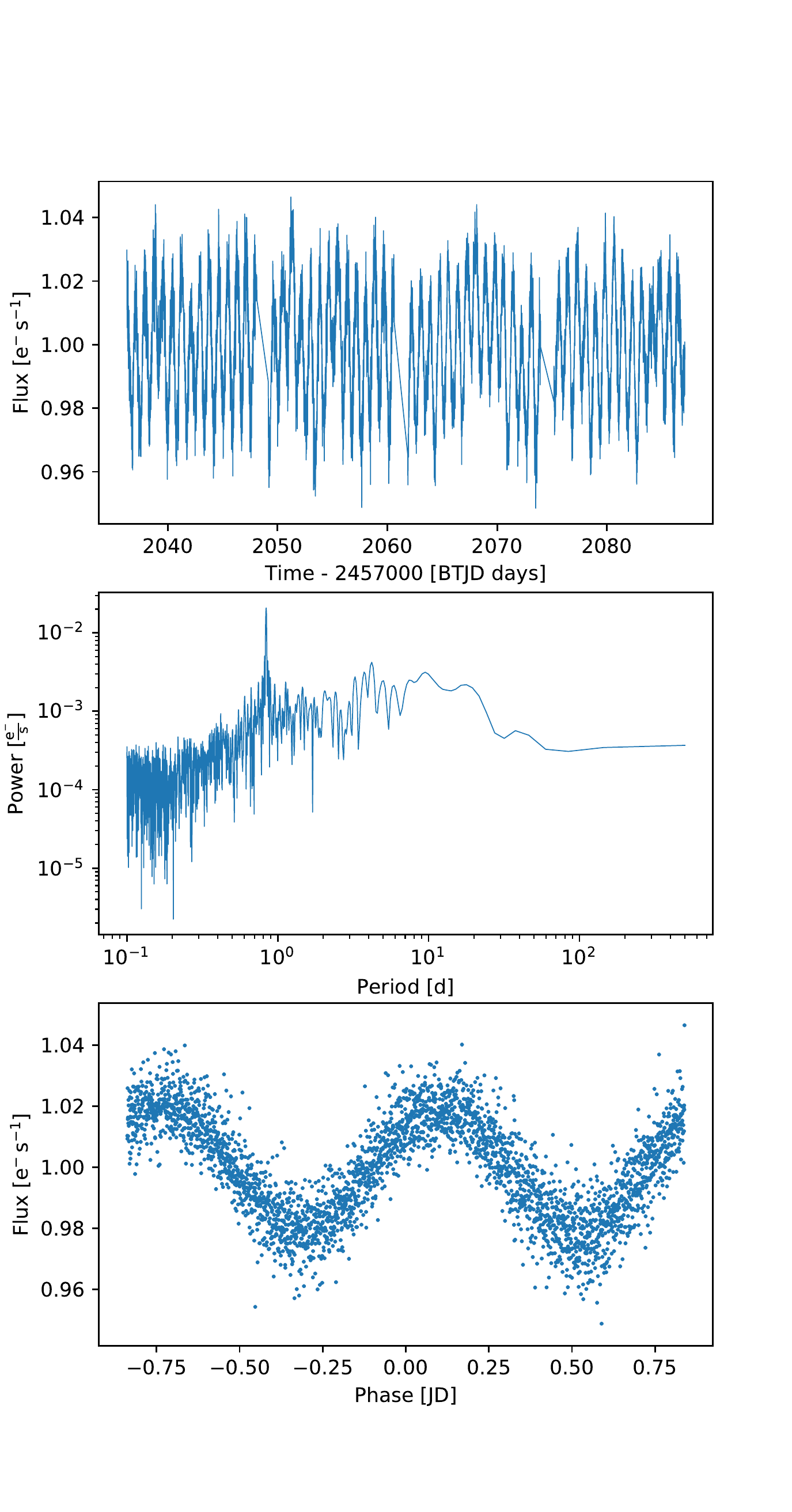}
      \caption{TESS light curves for \sgs\, extracted from FFIs taken in sectors 27 and 28. The middle panels show the periodograms of the light curve, and the bottom panel includes the phase-folded light curve with a period of 0.838 days.}
         \label{LC}
\end{figure}

\sgs\, was observed by the TESS space telescope in 2018 (sectors 1 and 2), 2019 (sector 13), and 2020 (sectors 27 and 28). We created our own light curves from the full frame images (FFIs) by
using the Lightkurve  package (Lightkurve Collaboration et al. 2018) to download a target pixel file with a 5 $\times$ 5 pixel image with 21 arcsec the projected pixel size, centered on the target for every available TESS sector. Based on cross-correlation with the Gaia DR3 catalog, the region surrounding the star is relatively sparse, with only one bright source one arcmin away. To exclude the contribution from this bright source, we opted for a one-pixel aperture mask centered on the source and further applied background subtraction. The Lightkurve package is also used to remove outliers and to detrend and normalize the light curve. The light curve is shown in Fig.\ref{LC} (upper panel); a periodic signal is immediately evident in the light curve and we measure a period of P =0.838 days. The ASAS-SN Variable Stars Database classified the star as an eclipsing binary with a period of 1.675 days based on its photometric variability \citep{Jayasinghe2020}; this is twice the period observed using the TESS light curve. They reported similar variability to W Ursae Majoris-type binary (EW-type), which are known as contact eclipsing binaries. EW binary light curves are characterized by rounded peaks and sharp minima, as well as an overall symmetric shape, and typically have periods of less than one day. Two components in this subclass are expected to have roughly the same temperature, making their primary and secondary minima nearly indistinguishable from one another. Despite the light curve features, the spectra indicate that the two components have different temperatures and properties than EW-type. 
The inflated stripped star and Be star are just too big to fit into such a short-period system. In addition, three spectra taken within one day only show an RV variability of $\lesssim10$\,km\,s$^{-1}$.

Rotational variability appears to be the most common type of light variation among B/Be stars
 (Balona et al.2016, 2019, 2021). Therefore, one possibility is that the observed variability is a result of the fast rotation of the secondary Be star. By plugging in the estimated radius of secondary from spectral analysis and the period from the light curve into  $\varv_\mathrm{rot}=2\pi R/ P_\mathrm{rot}$,  we found  $\varv_\mathrm{rot} \sim 480-620$\,km\,s$^{-1}$ or $i\sim 35-60\degr$. This range of inclination is in agreement with the shape of observed \ha and Brackett series emission line profiles.
 
\section{Additional plots}

\begin{figure*}[!htb]  
\caption{Observed spectra and photometry of  \sgs\, (blue) compared to the model SED and synthetic spectra.  The composite
model (dashed red) is the weighted sum of the stripped star primary (dotted brown)
and rapidly rotating B star secondary (dashed green) model spectra}
\label{fig:optfit}
\vspace{0.5cm}
\includegraphics[scale=0.9, angle =90]{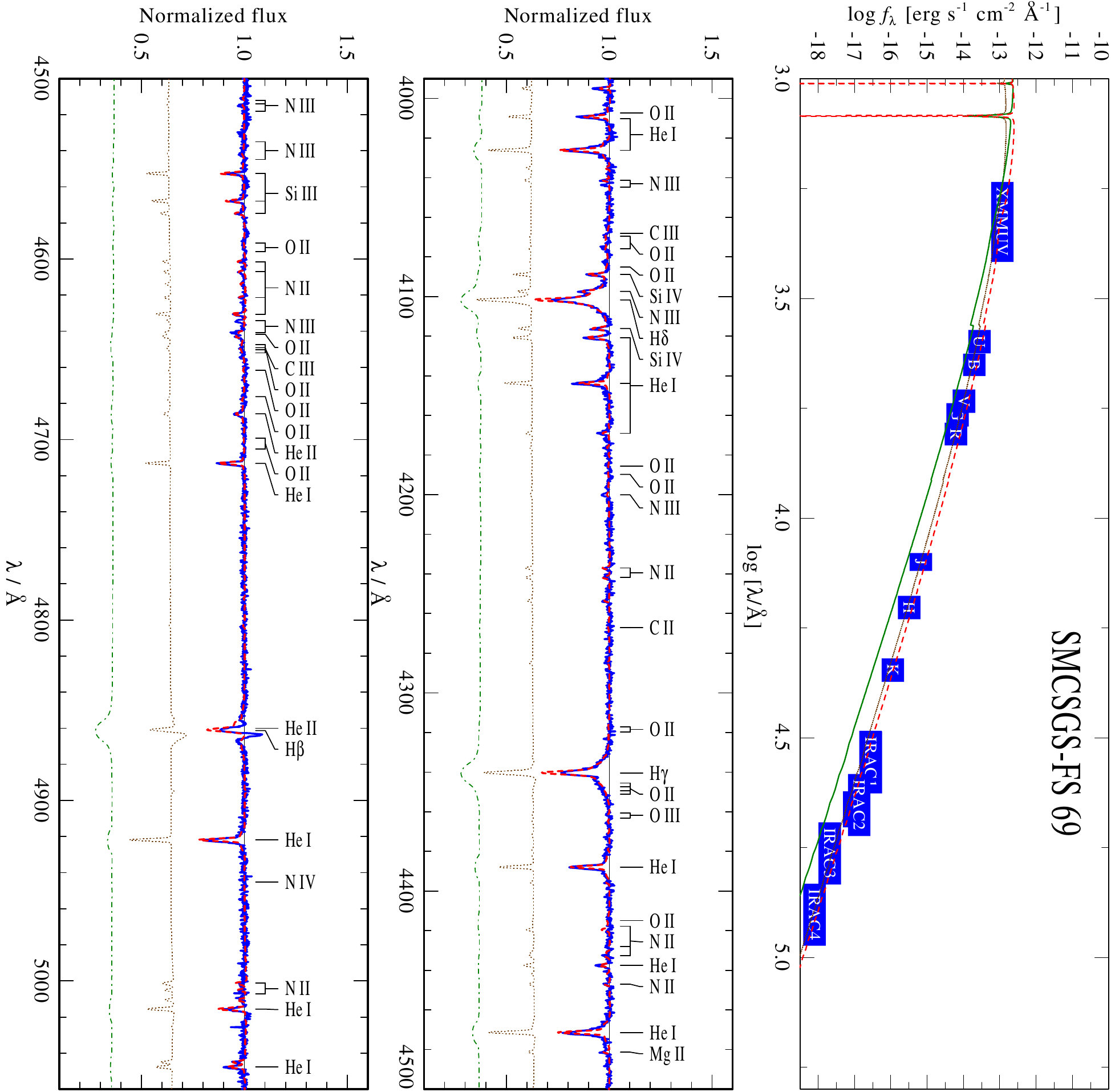}
\end{figure*}   

\begin{figure*}[!htb]  
\caption{Comparison of observed \ha profile (blue) with composite PoWR models. (Left) \ha reproduced by a model with strong wind ($\log \dot{M} =-6.2$) assuming it is mostly coming from the stripped star. (Right) \ha feature  reproduced by a composite (red dashed) of strong disk emission from the Be star (green dashed) and weak absorption component ($\log \dot{M} =-7.2$) from the stripped star (brown dotted). }

\label{fig:optfitha}
\vspace{0.5cm}
\includegraphics[scale=0.75,trim={8cm 0cm 0cm 0cm},clip]{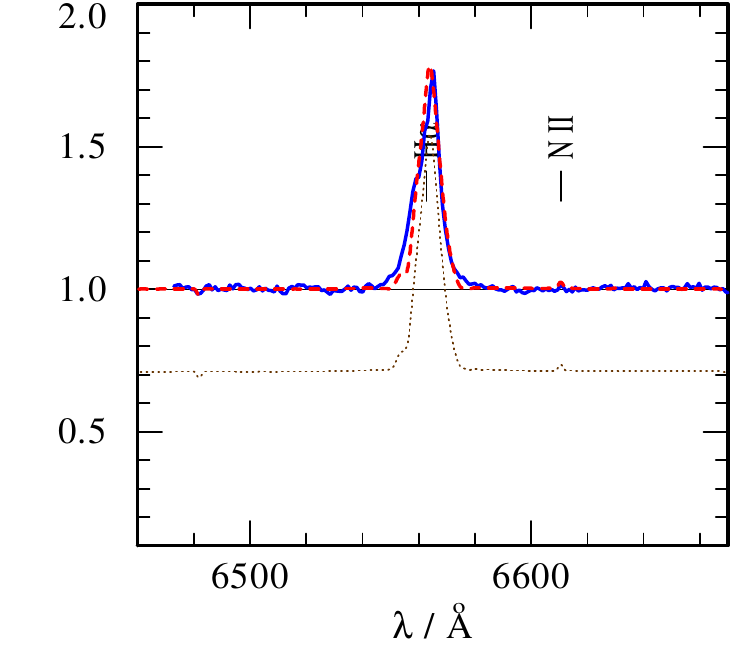}
\includegraphics[scale=0.75,trim={8cm 0cm 0cm 0cm},clip]{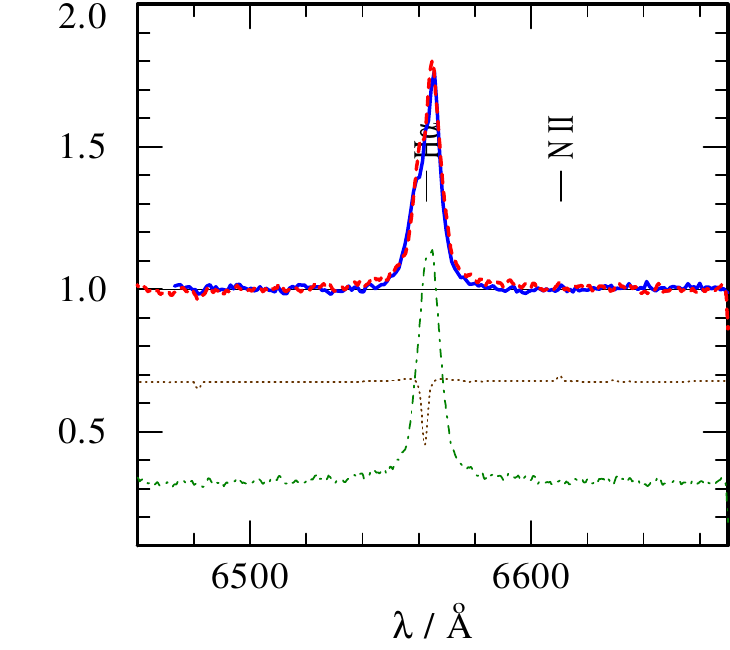}
\end{figure*}

\begin{figure*}[!htb]  
\caption{APOGEE H-band spectra of \sgs\, showing disk emission. The Brackett series lines are labeled}
\label{fig:IR}
\vspace{0.5cm}
\includegraphics[width=0.92\textwidth]{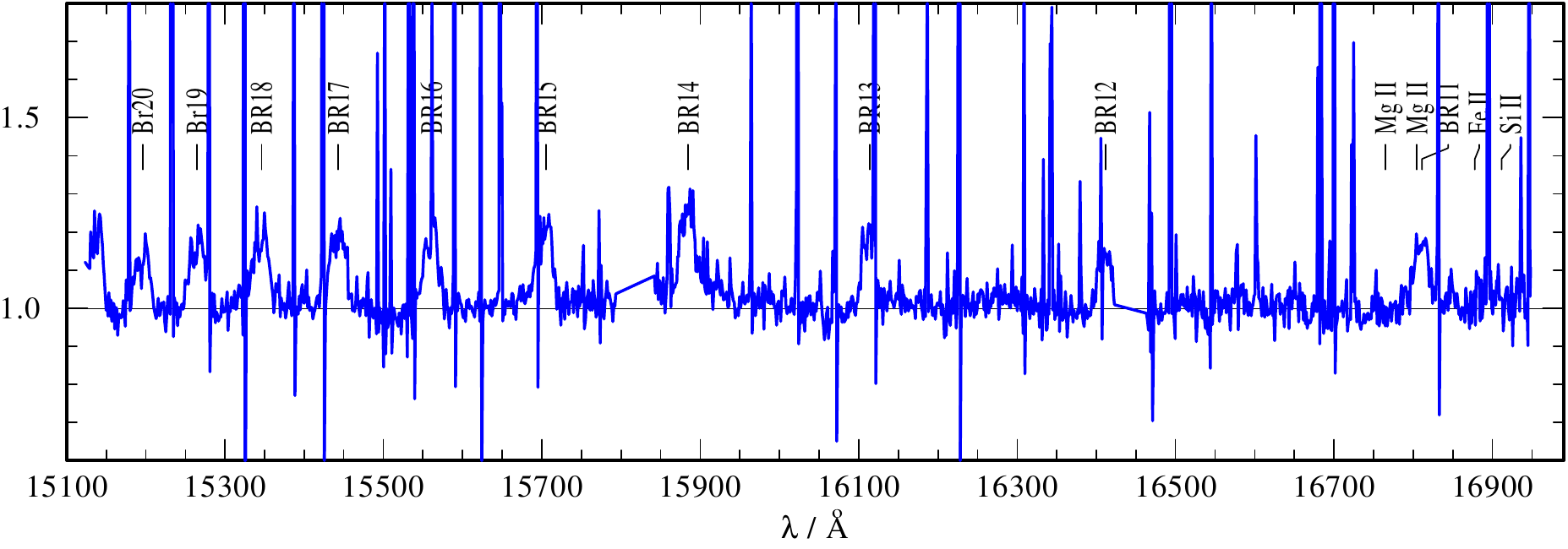}
\end{figure*}   

\section{Additional tables}

\begin{table*}
\centering
\caption{Absolute abundances in units of $12 + \log$\,(X$_{i}/$H) for the stripped star compared to SMC B main sequence and supergiants}
\label{table:abundance}
\setlength{\tabcolsep}{5pt}
\begin{tabular}{ccccccccc}
\hline 
\hline
\noalign{\vspace{1mm}}

& C & N & O & Si & Mg \\
\noalign{\vspace{1mm}}
\hline 
\noalign{\vspace{1mm}}
SMCSGS-FS\,69  & $\lesssim$6.15 & 8.16 & $\lesssim$6.87 & 6.82 & 7.12\\
SMC B supergiants\tablefootmark{1}& 7.30 & 7.55& 8.11 &6.75 &6.8\\
SMC B stars \tablefootmark{2}  &  7.23 (7.35)& 7.17 (6.50) & 8.06 (8.05) & 6.79 (6.80) & 6.74 (6.75)  \\
\hline
\end{tabular}
\tablefoot{
\tablefoottext{1}{average  abundances of SMC B supergiants adopted from \citet{Dufton2005}}
\tablefoottext{2}{average (and baseline) abundances of SMC B stars adopted from \citet{Hunter2007}}
}
\end{table*}

\section{Binary evolution modeling}

\label{sec:app_mesa}
The binary evolution modeling was carried out with the MESA\footnote{MESA version r15140, \url{http://mesa.sourceforge.net/}} stellar evolution code \citep{Paxton2015,Paxton2018,Paxton2019}. Most of the modeling assumptions follow those of \citet{Klencki2022} for the SMC composition. A notable difference is the use of the convective premixing scheme \citep[introduced in][]{Paxton2019}, which effectively introduces highly effective mixing in semiconvective regions. We model both binary components and finish the evolution once the primary reaches central carbon depletion. In most of our models we do not include stellar rotation in order to retain the option of choosing the accretion efficiency as a free parameter.\footnote{In MESA, a rotating secondary is unable to accrete any more mass once it is spun up to breakup velocity, which self-regulates the accretion rate during mass transfer.}
We explore the parameter range of primary masses from $10\,M_{\odot}$ to $18\,M_{\odot}$ (in steps of $0.2\,M_{\odot}$ around the best-fit model), mass ratios from $0.6$ to $0.97$, and orbital periods covering the entire range in which mass transfer interaction may occur. We vary mass accretion efficiencies from $0\%$ to $70\%$ in steps of $10\%$. 

To find the best-fitting model, we first try matching the surface properties of both stars derived in spectral analysis (Table \ref{table:parameters}), namely: the luminosities, effective temperatures, and surface gravities for both stars, as well as the surface abundances of H, He, C, N, and O for the primary. None of our binary models was able to reproduce all the stripped star properties simultaneously.

\begin{figure*}
   \centering
   \includegraphics[width=\textwidth]{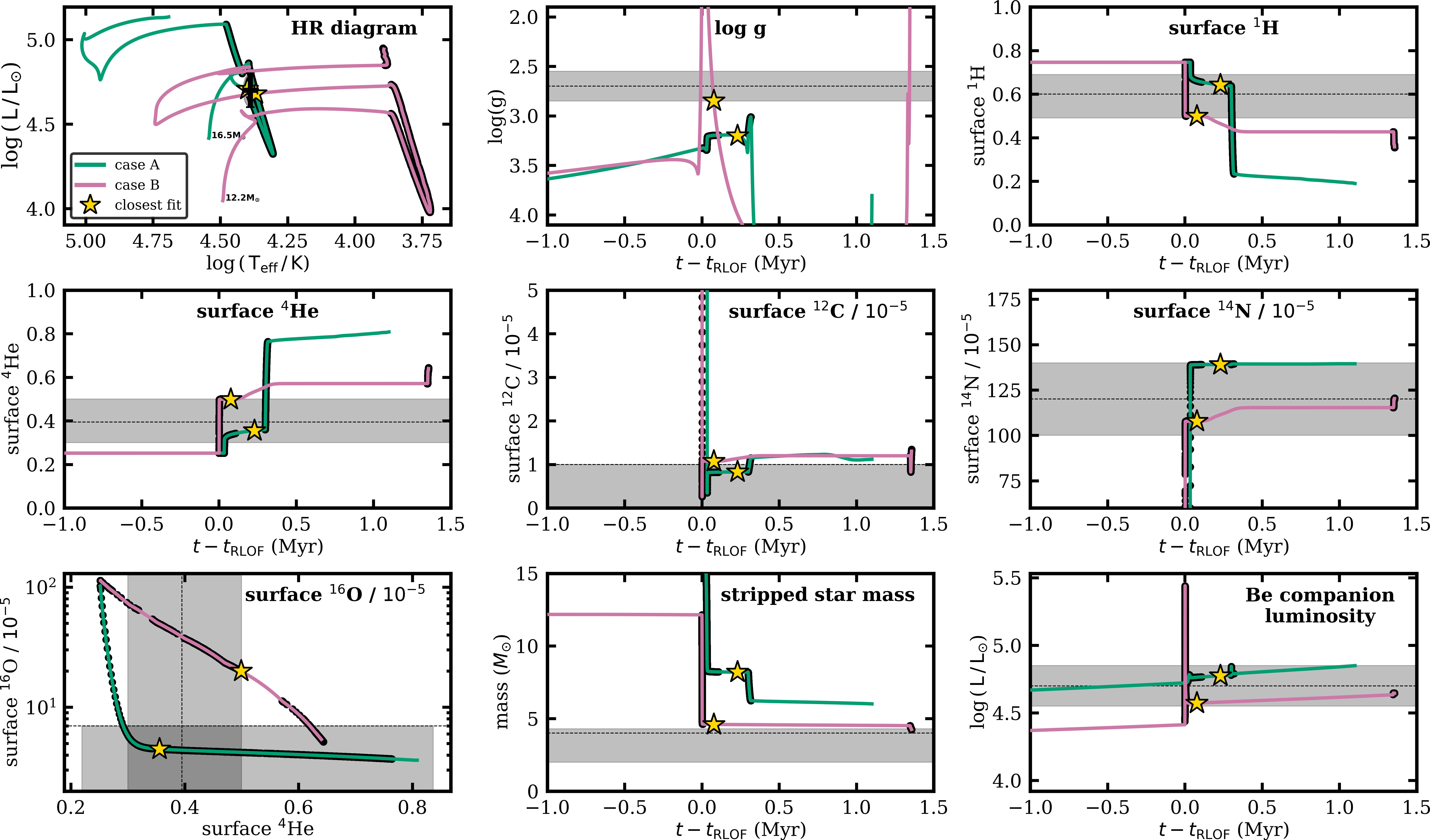}
      \caption{Stellar parameters and abundances derived spectroscopically (in grey) compared to binary evolutionary  model predictions, variants with Case A and Case B mass transfer evolution (see also Fig.\,\ref{hrd_bin}). The Case A mass travel model finds better agreement with surface abundances, although it is inconsistent with the measured surface gravity (and correspondingly, the inferred spectroscopic mass of the stripped star).}
         \label{fig:param}
\end{figure*}

Models in which the stripped star was produced following a Case A + Case AB mass transfer evolution were characterized by a surface He abundance $\gtrsim 0.75$ (in tension with the measured $0.3-0.5$ range). Models of stripped stars produced via Case B mass transfer were generally in better agreement with the measurements. However, we could only marginally reproduce the surface He and H abundances, with most models predicting $X_{\rm He} > 0.5$ and $X_{\rm H} < 0.5$ for the (partially) stripped star stage. Marginal agreement was found in those models where a relatively large amount of envelope was left during mass transfer ($\sim 1 M_{\odot}$ in  
our favorite Case B model), which would preferentially occur in wider binaries (periods of hundreds rather than tens of days). 
The requirement of $X_{\rm He} < 0.5$ and $X_{\rm H} > 0.5$ further prevented us from finding a matching model for the surface O abundance. We found that stripped stars produced in Case B mass transfer would gradually reduce their surface O with increasing surface He abundance, as illustrated by the Case B example in the lower left panel of Fig.\,\ref{fig:param}. However, the exact shape of the relation between the He and the O abundance of the inner envelope layers is subject to large uncertainties related to the extent of internal mixing above the H-burning shell through processes such as overshooting, semiconvection, and rotationally induced mixing. The difficulty in simultaneously reproducing the H, He, and O abundance of the stripped star in \sgs\, is likely a reflection of the limitation of current stellar models in predicting the chemical structure of stars in this complicated inner region.

Altogether, we find that under the assumption of Case B mass transfer evolution, the observed properties of the stripped star are best matched with progenitors of $\sim 12.2 M_{\odot}$.
In reproducing the properties of the secondary, there is a level of degeneracy between the initial mass ratio and the assumed accretion efficiency: the more unequal the initial mass ratio, the higher the accretion efficiency needed to reproduce the current luminosity of the secondary.\footnote{In principle, additional constraints are offered by the effective temperature of the Be companion. An 8 $M_{\odot}$ secondary that accretes 7 $M_{\odot}$ of material would be less evolved and therefore hotter than a 12 $M_{\odot}$ secondary that accretes $3 M_{\odot}$ of material. However, since most of our models do not include the fast rotation of the Be star, we most likely somewhat overestimate the temperature of the secondary.} A particularly interesting solution is a case when the binary is formed with the initial mass ratio very close to unity. In such a system, the amount of mass that is needed to be accreted by the secondary in order to explain its current luminosity can be viewed as the lower limit, placing constraints on the minimum accretion efficiency required to reproduce the observed binary.

\begin{figure}
   \centering
   \includegraphics[width=\columnwidth]{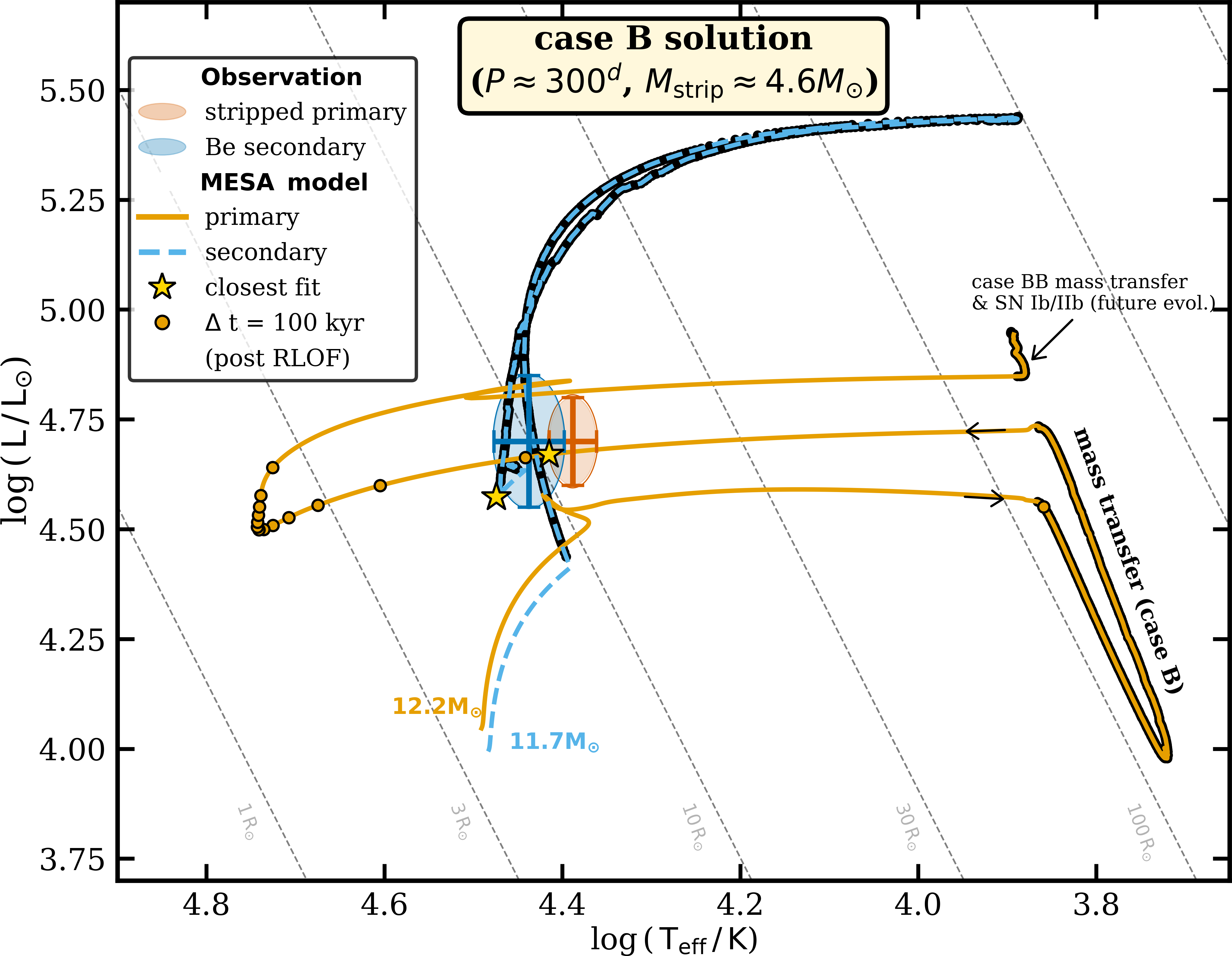}
      \caption{Best-fitting binary evolution model in which the stripped star is produced as a result of Case B mass transfer. The model requires accretion efficiency of at least $\sim 40\%$ to explain the current luminosity of the Be star secondary. }
         \label{fig:hrd_caseB}
\end{figure}

This scenario is found in our favorite Case B binary model, shown in Fig.\,\ref{fig:hrd_caseB}, where the system is formed as a $12.2 M_{\odot}$ primary and an $11.7 M_{\odot}$ secondary on a 126-day period orbit. The secondary accretes with $\sim 40\%$ efficiency, and once it regains thermal equilibrium at the end of the mass transfer phase it settles back onto the MS with a luminosity of $\log L/L_\odot\!\sim\!4.5$  (i.e., on the lower end of the range measured for the Be star in \sgs).

As an alternative scenario, we calculated a subgrid of models that include rotation, limiting the amount of material that can be accreted by the secondary. This has severe implications for the Case B models, as it prevents the secondary from accreting enough material to match the observed parameters. However, we were able to find an acceptable solution with a model undergoing Case A mass transfer. This model allows for higher initial primary masses, and hence also higher initial secondary masses without the need for the secondary to accrete a significant amount of material.

\begin{figure}
   \centering
   \includegraphics[width=\columnwidth]{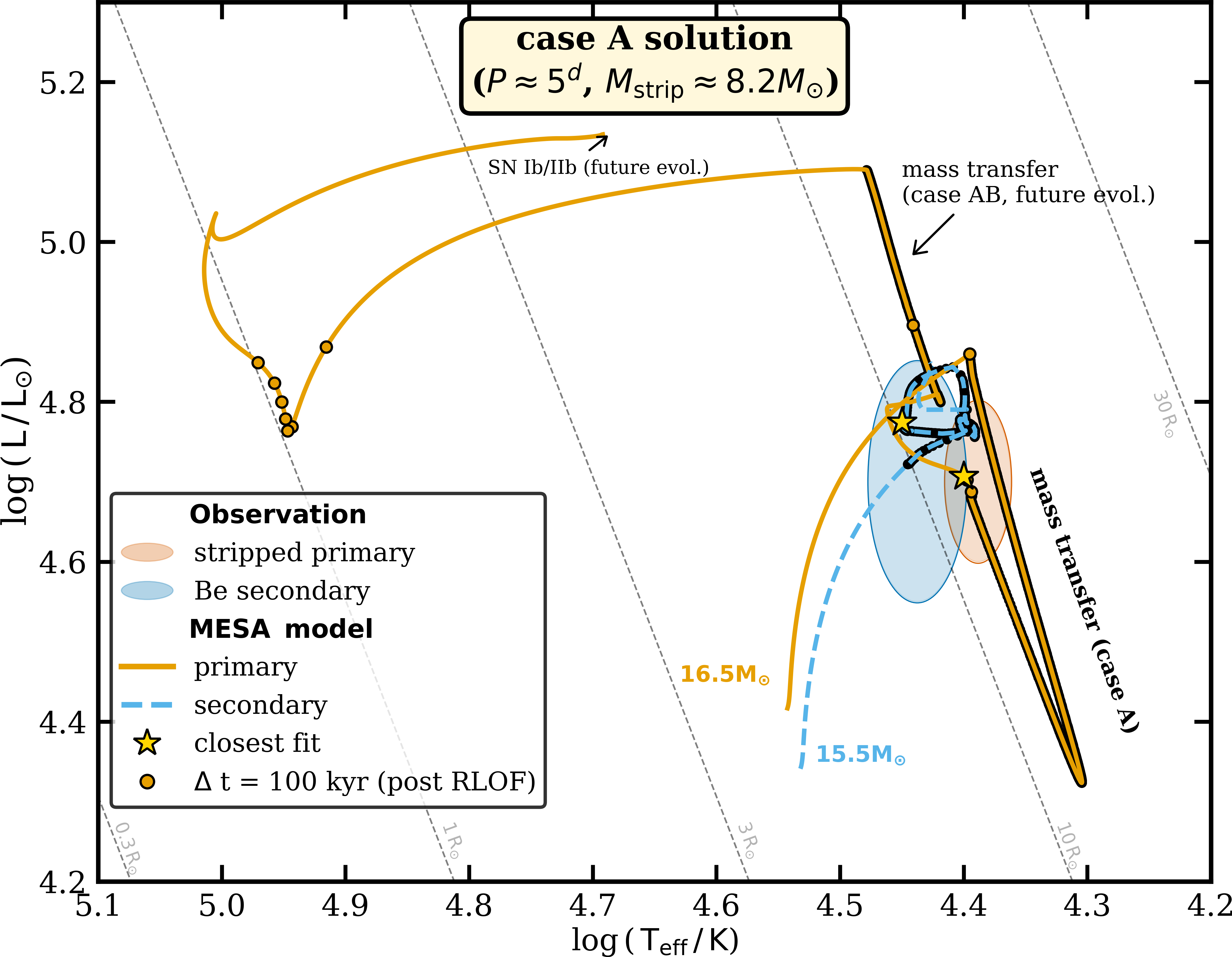}
      \caption{Best-fitting binary evolution model in which the stripped star is produced as a result of Case A mass transfer and is currently core-H burning.  }
         \label{fig:hrd_caseA}
\end{figure}

Our favorite case~A model has an initial primary mass of $16.5\,M_\odot$, an initial secondary mass of $15.5\,M_\odot$, and an initial orbital period of only $4.85\,\mathrm{d}$. The corresponding stellar evolutionary tracks are depicted in Fig.\,\ref{fig:hrd_caseA}. In this model the primary initiates mass transfer close to the terminal-age main-sequence (i.e., shortly before core-H exhaustion). During the mass transfer the secondary only accretes about $\lesssim1\,M_\odot$ and quickly reaches critical rotation, preventing it from accreting more material. With this model we are able to explain the derived rotation rates of the two binary components, but also the observed surface abundances of H, He, C, N, and O simultaneously (see Fig.\,\ref{fig:param}). Despite the good agreement of the surface properties of the primary component, we are unable to explain the observed low spectroscopic mass and require the primary to be twice as massive as derived. It is worth mentioning, that the envelope mass of the primary in this case makes up roughly 50\% of its total mass.

For the future evolution of the system our favorite Case A model predicts that the primary will expand again after all H is exhausted in the core. This will coincide with a new, rapid mass-transfer episode called Case AB, during which most of the H-rich envelope will be removed. The primary is predicted to evolve quickly into a He-star that still has a thin ($\lesssim0.5\,M_\odot$) H-poor ($X_\mathrm{H}\sim0.2$) envelope. Similar to the Case B model, the primary will expand again during He-shell burning, and will eventually explode as a supernova, most likely of type IIb/Ib. If the binary survives the explosion, the system will further evolve into a high-mass X-ray binary with a NS accretor. The relatively short-period of the X-ray binary predicted in the Case A scenario (from a few days to a few tens of days) means that in the event of a common-envelope the binary would likely merge and produce a Thorne-Żytkow object. 

To estimate the relative rate of the two scenarios, we consider that the case A solution originates from binaries with initial periods  between $\sim$2 and $\sim$5 days and initial primary masses between 16 and $17\,M_\odot$. Such systems will interact during the later half of the MS evolution of the primary \citep[at SMC metallicity of $0.2\,Z_\odot$,][]{Klencki2020}, making it possible for the detached donor to match the temperature and luminosity of the stripped star in \sgs. Similarly, we consider that the case B scenario occurs for binaries with initial primary masses between 11.5 and $12.5\,M_\odot$ and initial periods between 5 and 1000 days (the period range for case B mass transfer evolution in the considered mass range). Based on that, assuming the initial mass function of \cite{Kroupa2013} and the initial period distribution for massive binaries from \cite{Sana2012}, we estimate that the formation rate of the progenitors of the case B scenario is about six times higher than that of the case A scenario. On the other hand, the lifetime of the stripped-star stage consistent with \sgs\, is typically about five times longer in the case A scenario ($\sim$100-200 thousand years) compared to the case B scenario (30 thousand years). Altogether, this crude estimate shows that the relative rate of occurrence of partially stripped stars similar to \sgs\, is comparable for the two scenarios.

 
\end{document}